\documentclass{PoS}

\title{Addressing $R_K$ and neutrino mixing in a class of $U(1)_X$ models}

\ShortTitle{$R_K$ anomaly and neutrino mixings with $U(1)_X$}

\author{\speaker{Disha Bhatia}\thanks{I would like to thank the organizers for giving me the opportunity to present our work at the 9th International Workshop on the CKM Unitarity Triangle.}\\
        Tata Institute of Fundamental Research, Mumbai 400005, India\\
        E-mail: \email{disha@theory.tifr.res.in}}

\author{Sabyasachi Chakraborty\\
        Tata Institute of Fundamental Research, Mumbai 400005, India\\
        E-mail: \email{sabya@theory.tifr.res.in}}

\author{Amol Dighe\\
        Tata Institute of Fundamental Research, Mumbai 400005, India\\
        E-mail: \email{amol@theory.tifr.res.in}}

\abstract{We present a class of minimal $U(1)_X$ models as a plausible solution to
          the $R_K$ anomaly that can also help reproduce the neutrino mixing pattern.
          The symmetries and the corresponding $X$-charges 
          of the fields are determined in a bottom-up 
          approach demanding both theoretical and experimental consistencies.
          The breaking of $U(1)_X$ symmetry results in a massive $Z^{\prime}$, whose 
          couplings with leptons and  quarks 
          are necessarily non-universal to address the $R_K$ anomaly.
          In the process, an additional Higgs doublet is introduced to generate quark mixings.
          The mixings in the neutrino sector are generated through Type-I seesaw mechanism by the addition of 
          three right handed neutrinos and a scalar singlet. 
          The $Z^{\prime}$ can be probed with a few
          hundred fb$^{-1}$ of integrated luminosity at the 13 TeV LHC in the di-muon channel.                           
          }

\FullConference{9th International Workshop on the CKM Unitarity Triangle\\
		28  November - 3 December 2016\\
		Tata Institute for Fundamental Research (TIFR), Mumbai, India}

\begin{document}

\section{Introduction}
The flavour observable, $R_K \equiv {\rm BR}( B \to K \mu \mu) /{\rm BR}( B \to K e e) $, 
is predicted to be unity in the standard model (SM) to a very good accuracy~\cite{rk_sm_qcd,rk_sm_qed}.
The experimentally measured value of $R_K$ in the low dilepton mass squared bin~\cite{RKexp} 
is $0.745^{+0.090}_{-0.074}\pm 0.036$, which deviates from the SM currently by 2.6$\sigma$ and 
therefore hints towards lepton flavour universality violation. 
The angular observable $P_5^\prime$~\cite{DescotesGenon:2012zf} measured in $B \to K^* \mu \mu$ also 
shows a deviation from the SM~\cite{P52,Wehle:2016yoi}. 
The $b \to s$ flavour anomalies in $R_K$ and $P_5^\prime$
can be simultaneously addressed if the new physics (NP) effects are present 
in the Wilson coefficients ($C_i$) of the following operators ($\mathcal{O}_i$)~\cite{rk_eft}: 
\begin{eqnarray}
\mathcal{O}_9^\ell &=& \left(\overline{b}\, \gamma_\mu P_L \,s \right)
         \,\,\left( \overline{\ell} \,\gamma^\mu \ell\right)\;, \quad 
\mathcal{O}_{10}^\ell \,\,=\,\, \left(\overline{b}\, \gamma_\mu P_{L}\,s\right)
\,\, \left(\overline{\ell} \,\gamma^\mu \gamma_5 \ell\right)\;, \nonumber \\
\mathcal{O}_9^{\ell{\prime}} &=& \left(\overline{b}\, \gamma_\mu P_R\,s\right)
 \,\, \left(\overline{\ell} \,\gamma^\mu \ell\right) \;,\quad 
\mathcal{O}^{\ell{\prime}}_{10} \,\,=\, \,\left(\overline{b}\, \gamma_\mu P_{R}\,s\right)
 \,\, \left(\overline{\ell}\, \gamma^\mu \gamma_5 \ell\right)\;.
\end{eqnarray}

Global fits~\cite{Descotes-Genon:2013wba,globalfit1,globalfit2,globalfit3,globalfit4} performed on $b \to s \ell \ell$ and $b \to s \gamma$ data 
prefer dominant NP effects in $\mathcal{O}_9^{\mu}$, with
$C_9^{{\rm NP},\mu} \sim -1$.
Non-zero contributions to $C_9^{{\rm NP},e}$ and $C_9^{{\rm NP},\mu}$
simultaneously are consistent with global fits~\cite{globalfit1,globalfit2,globalfit3,globalfit4} within 2$\sigma$.
The two-dimensional fit performed by~\cite{globalfit4} favour NP effects in
$\left(C_9^{ \mu},C_9^{ e}\right)$ over others. 

Motivated by above results, we construct our model in a bottom-up approach 
where NP contributions are present in the $\mathcal{O}_9^\ell$ operator~\cite{ourpaper}.
New physics contributions to $C_9^\ell$ can be generated in the presence of a $Z^\prime$ 
or leptoquarks. We choose to explain the anomaly using a $Z^\prime$ solution, and 
hence augment the SM with an additional gauge symmetry, $U(1)_X$~\cite{ourpaper}.
The SM fields
$i$ are assigned the charge $X_i$ under $U(1)_X$. The explanation 
of the $R_K$ anomaly necessarily requires the $X$-charges of the electron and muon 
to be different. 
Dominant NP contributions to $b \to s \ell \ell$ using a $Z^\prime$ also dictate 
non-universality of the $X$-charges for quarks.

The $U(1)_X$ extensions of the SM should also be able to explain neutrino masses and mixings. Thus 
it will be interesting to have a common origin for flavour anomalies and neutrino mass generation.
We do not pre-assume any value for the $X$-charges, rather determine 
them in a bottom-up approach while satisfying all the current measurements and 
theoretical consistencies in a minimalist way. 
This also provides a framework which can be used for analyzing future data.

\section{Constructing the $U(1)_X$ models}

\subsection{Theoretical considerations}
The additional gauge symmetry, $U(1)_X$ should not introduce any gauge anomaly. 
A minimal addition of three right handed neutrinos 
with vector-like $X$-charge assignments, i.e.,
$X_{u_{L}} =  X_{d_{L}} =  X_{u_{R}} =  X_{d_{R}} 
\equiv X_Q \;,
 X_{\ell_{L}} =  X_{{\nu_\ell}_{L}} 
=  X_{\ell_{R}} = X_{{\nu_\ell}_{R} } \equiv  X_\ell  \;,$
will ensure our model to be gauge anomaly free, 
if the charges satisfy
\begin{eqnarray}
\sum_i 3 \,X_{Q_i} + X_{\ell_i} = 0 \;,
\label{eq:anomaly}
\end{eqnarray}
where the sum is over the fermionic generation. 
With these charges, the Yuwaka interactions, given as
\begin{equation}
\mathcal{L}_{\rm Yuk} =  \overline{{Q_{L}}}\, \mathcal{Y}^d \Phi \,d_R 
                    +    \overline{Q_L} \,\mathcal{Y}^u \Phi\,u_R 
                    +    \overline{L_L} \,\mathcal{Y}^e  \Phi\,e_R\;,
\end{equation}
will consist of non-zero diagonal $\mathcal{Y}^d_{ii}$,
$\mathcal{Y}^u_{ii}$ and $\mathcal{Y}^e_{ii}$ only if $\Phi_{\rm SM}$ is a singlet under 
$U(1)_X$.

\subsection{Constraints on the quark sector from the SM predictions}
The unequal $X$-charges for the quark generation leads to 
potential flavour changing neutral interactions at tree level, 
which can affect neutral meson mixings, in particular. The stringent
constraints from $K$--$\overline{K}$ mixing~\cite{utfit} are
accounted for by choosing $X_{Q_1} = X_{Q_2}$. 
Hence in our model, the non-universality appears in the third generation
of the quark sector. Although this will generate appropriate NP contributions
to explain $R_K$, it will be unable to generate the measured CKM matrix. 
This problem can be resolved with an additional doublet, $\Phi_{\rm NP}$, with $X$-charge, 
$d = X_{Q_1} - X_{Q_3}$~\cite{ourpaper} (A similar choice has been considered in~\cite{horizontalpaper}).
We also choose the left-handed quark rotation matrices 
$V_{d_{L}} = V_{\rm CKM} \;{\rm and}\; V_{u_L} = I \;,$
which ensure no $Z^\prime$ contribution to the CP violating phases 
in $B$--$\overline{B}$ oscillations. With this choice, 
$[V_{d_R}]_{23}  \approx A \lambda^2 m_s/m_b \;,  [V_{d_{R}}]_{13} \approx - A \lambda^3 m_d/m_b
\;,  [V_{u_R}]_{23} = [V_{u_R}]_{13} = 0 \;,$
with unconstrained  $[V_{d_R}]_{12}$ and $[V_{u_R}]_{12}$, which we choose to be vanishing. 
Thus, $V_{d_{R}} \approx I$ and $V_{u_R} = I$ in our model.  

Following the results from global fits~\cite{globalfit1,globalfit2,globalfit3,globalfit4},
we require dominant new physics contributions only to $\mathcal{O}_9$. Therefore,
the contributions to other operators --- $\mathcal{O}_9^\prime$,
$\mathcal{O}_{10}$ and $\mathcal{O}_{10}^\prime$ should vanish. The vector-like charge assignments
automatically ensure the NP contributions to $\mathcal{O}_{10}$ and $\mathcal{O}_{10}^\prime$ are absent. 
The contributions to the $\mathcal{O}_9^\prime$ operator are also small in comparison to the $\mathcal{O}_{9}$
as $V_{d_R} \approx I$. Therefore our charge assignments generate significant NP contributions only to 
the $\mathcal{O}_9$ operator~\cite{ourpaper}.



\subsection{Neutrino masses and mixings}

We generate neutrino masses and mixings using Type-I seesaw mechanism.
The additional scalar $S$, with $X$-charge $a$, is included to obtain the observed neutrino
mixings from the oscillation data. The Lagrangian describing the mass term of neutrinos 
(with $X$-charges $y_e$, $y_\mu$ and $y_\tau$) is

\begin{equation}
\mathcal{L}^{\rm mass}_{\nu} = -\overline{\nu_L} \,m_D\, \nu_R -\frac{1}{2} \overline{\nu^c_R} \,M_R\, \nu_R  
-\frac{1}{2} \overline{\nu^c_{R}} {\mathcal Y}_{R} \nu_{R} \,S + h.c. \;.
\label{eq:L-mass-nu}
\end{equation}
Without loss of generality, we can always choose the basis for the charged leptons
and left handed neutrinos $v_L$ such that $m_D$ and $m_\ell$ (mass matrix for charged leptons) 
are diagonal. The $X$-charges dictate the texture of the right handed mass matrix, $M_R^S$: 
\begin{equation}
  {[M^{S}_R]}_{\alpha\beta} = {[M_R]}_{\alpha\beta} + \frac{v_S}{\sqrt{2}}[y_{R}]_{\alpha\beta} 
  \neq 0 \quad {\rm if} \;\; y_\alpha + y_\beta = 0\;, \pm a \;.
  \label{eq:seesaw}
 \end{equation}
The allowed $M_R^S$ textures~\cite{2minors,1minors} can be used to infer the possible
$X$-charges in the lepton sector using eq.~(\ref{eq:seesaw}).
The allowed symmetry combinations hence obtained are listed in 
the left panel of fig.~\ref{fig:sel} and are clubbed 
according to $y_e/y_\mu$.
Note that the combinations obtained for the two-zero
textures match with those derived in~\cite{textures}.
\subsection{Scalar sector}
We have three scalars, $\Phi$, $\Phi_{\rm NP}$ and $S$, 
in our model, with $X$-charges $0$, $d$ and $a$, respectively. 
The doublet $\Phi_{\rm NP}$ and the scalar $S$
break $U(1)_X$ symmetry spontaneously, thereby providing $Z^\prime$ a mass.
The vacuum expectation value of $S$ is $\mathcal{O}({\rm TeV})$ owing to 
stringent limits on $Z^\prime$ mass from colliders. 
With such a vev, $S$ gets effectively decoupled and the effective scalar 
potential for the doublets, $\Phi \equiv \Phi_2$ and $\Phi_{\rm NP} \equiv \Phi_1$, is

\begin{eqnarray}
  V_{\Phi_1\Phi_2} &=&    - \left(m_{11}^2 -\frac{\lambda_{1S{_1}}}{2} v^2_{S}  \right) \Phi_1^\dagger\Phi_1 
           + \frac{\lambda_1}{2} (\Phi_1^\dagger \Phi_1)^2 
           - \left(m_{22}^2 - \frac{\lambda_{2{S}}}{2} v^2_{S}\right) \Phi_2^\dagger\Phi_2 \nonumber \\
          && + \frac{\lambda_2}{2} (\Phi_2^\dagger \Phi_2)^2 
             + \lambda_3 (\Phi_1^\dagger \Phi_1) (\Phi_2^\dagger \Phi_2) + \lambda_4 (\Phi_1^\dagger \Phi_2) (\Phi_2^\dagger \Phi_1) .
\end{eqnarray}

The absence of $\Phi_1^\dagger \Phi_2 + h.c.$ term in the Lagrangian would render a massless boson
in our theory. This problem can be avoided by equating $X_S = X_{\Phi_1}$, i.e. $a=d$, which admits
\begin{equation}
\Delta V_{\Phi_1\Phi_2 S} = -\widetilde{m}_{{12}} \left[ S \, \Phi_1^\dagger \Phi_2 
+ S^\dagger \, \Phi_2^\dagger \Phi_1\right] \;.
\end{equation}
This explicitly breaks the global $U(1)_A$ symmetry.
We now summarize the charge assignments in the following table:
\begin{table}[h!]
\begin{center}
\begin{tabular}{|c|c|c|c|c|c|c|c|c|c|}
\hline
 Fields & $Q_1$ & $Q_2$  & $Q_3$ & $L_1$ & $L_2$ & $L_3$ & $\Phi$ & $\Phi_{\rm NP}$ & $S$ \\ \hline
 $U(1)_X$ & $x_1$ & $x_1$ & $x_1-a$ & $ y_e$ & $y_\mu$ & $y_\tau$ & $0$  & $a$ & $a$\\ \hline
\end{tabular}
\end{center}
\caption{\label{tab:1} Vector-like $X$-charge assignments.
The fields $Q_i$ and $L_i$ refer to the $i^{\rm th}$ generations of quarks and leptons, respectively.
The charges of fermions are related with eq.~(\ref{eq:anomaly}). }
\end{table}


The $X$-charges of all particles are proportional to $a$~\cite{ourpaper},
therefore we absorb $a$ in the 
definition of $g_{Z^\prime}$ and proceed our analysis further by assigning $a=1$.
Note that, for simplicity, we continue to label the full $U(1)_X$ symmetry in terms of the leptons
as described in left panel of the fig.~{\ref{fig:sel}}.
\subsection{Selecting the plausible symmetry combinations}  
\begin{figure} [t]
\begin{minipage}[t]{0.3\linewidth}
\par\vspace{0pt}
\centering
\def\arraystretch{1.1}
\begin{tabular}{|c|c|}\hline
Category  & Symmetries\\ \hline
A & $L_\mu$, $L_\mu-L_\tau$\\ \hline
B   & $L_e - 3 L_\mu \pm L_\tau$\\ \hline
C   & $L_e + 3 L_\mu - L_\tau$\\ \hline
D  &$L_e -L_\mu \pm L_\tau$, $L_e - L_\mu \pm 3 L_\tau$\\ \hline
E & $L_e + L_\mu - L_\tau$, $L_e + L_\mu -3 L_\tau$ \\ \hline
F  & $3 L_e - L_\mu - L_\tau$\\ \hline
G   & $\,L_e$\\ \hline
\end{tabular}
\par\vspace{0pt}
\end{minipage}
\hfill
\begin{minipage}[t]{0.6\linewidth}
\par\vspace{0pt}
\centering
\includegraphics[height=5.5cm]{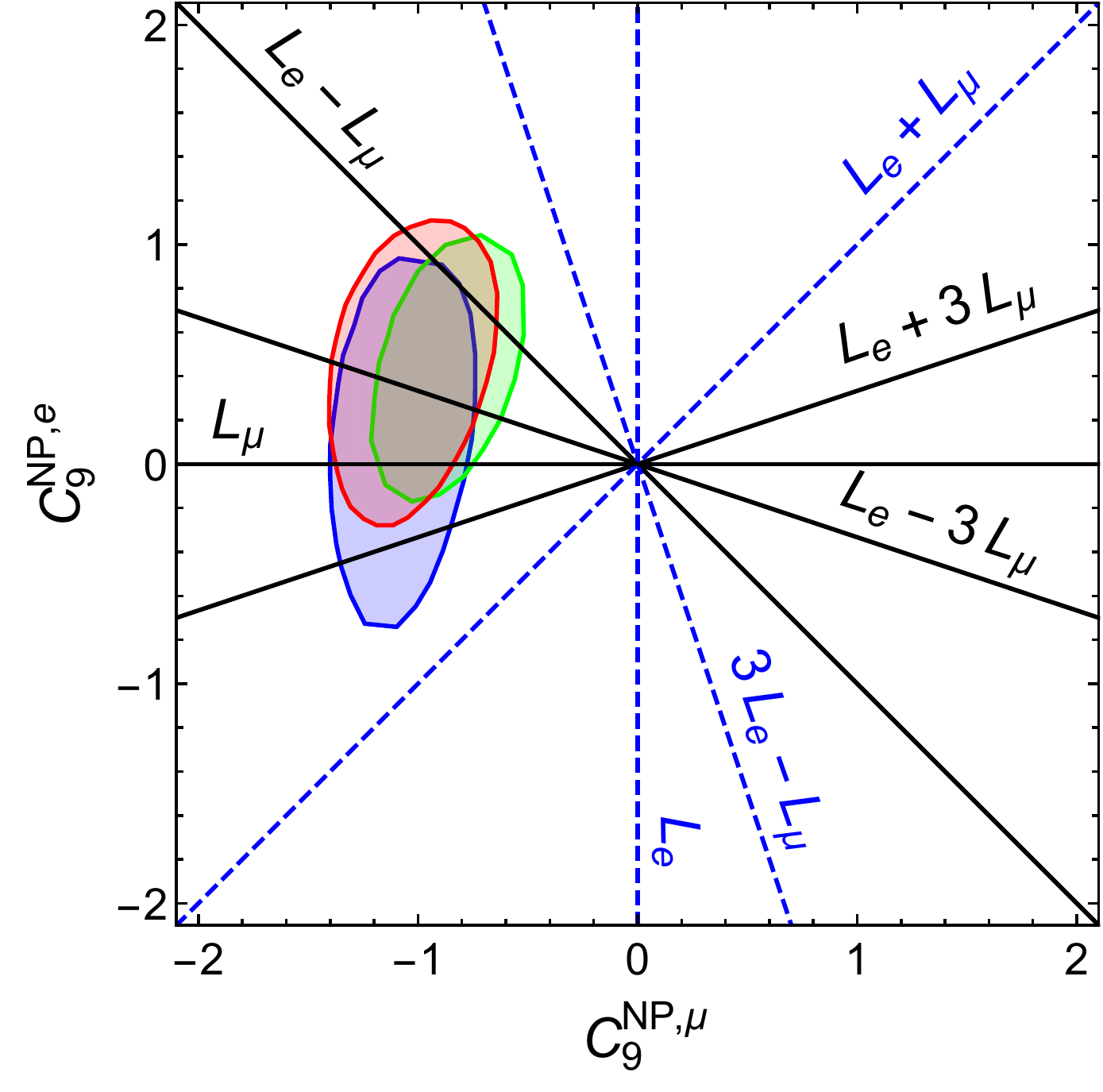}
\par\vspace{0pt}
\end{minipage}
\caption{{\label{fig:sel}} In the left panel, all the allowed leptonic symmetries consistent 
with the neutrino oscillation data are listed. They are categorized according to the ratio $y_e/y_\mu$.
In the right panel, the global fits obtained from~\cite{globalfit2} (red),~\cite{globalfit3} (blue) and~\cite{globalfit4} (green) are plotted together with the predictions for the symmetry combinations listed in the left panel, in the $\left( C_9^{{\rm NP},\mu}, C_9^{{\rm NP},e}\right)$  plane.}
\end{figure} 
We now want to 
determine the desirable symmetry combinations which are compatible
with the global fits to $b \to s \ell \ell$ and $b \to s \gamma$ data.
We consider the allowed region in the $\left( C_9^{{\rm NP},\mu}, C_9^{{\rm NP},e}\right)$
plane from global fits~\cite{globalfit2,globalfit3,globalfit4} and plot the predictions for all the symmetries listed 
in left panel of fig.~\ref{fig:sel}. The combinations in categories A, B, C and D
pass through $1\sigma$ contours for all the global fits and hence are selected~\cite{ourpaper}. 

\section{Experimental constraints}
\begin{figure} [t]
\begin{center} 
 \includegraphics[height=5.5cm]{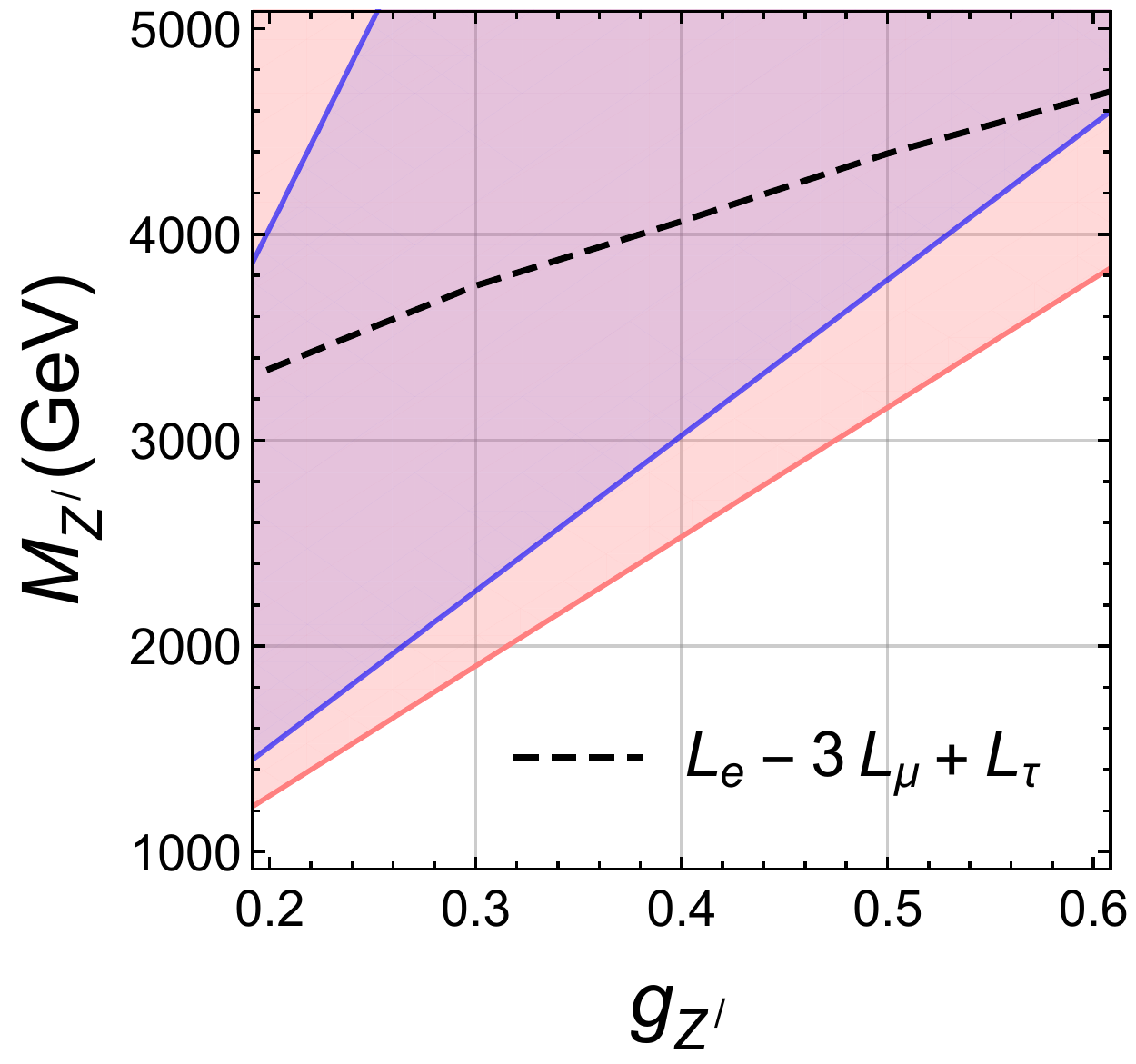}\quad
 \includegraphics[height=5.5cm]{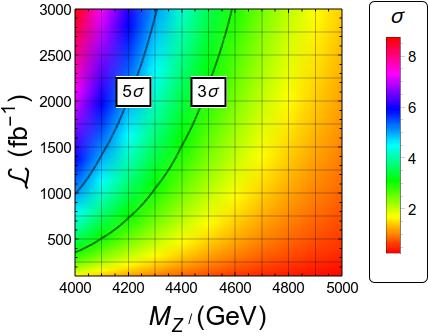}
 \caption{\label{fig:combinedconstraints}
  In the left panel, we plot the constraints obtained from neutral meson mixing, global fits on $b \to s \ell \ell$,
  $b \to s \gamma$ data and direct detection of $Z^\prime$ in the di-lepton channel. The regions in pink (blue) are 
  allowed by the neutral meson mixings (global fits) at $2\sigma$. The regions above the dotted line is consistent with direct detection limits at $95\%$ C.L.  In the right panel, we plot the significance of observing such a $Z^\prime$ (with $g_{Z^\prime}=0.36$) in the di-muon channel. The plots are shown for the symmetry combination $L_e-3L_\mu+L_\tau$.  
}
\end{center} 
\end{figure}

In this section, we consider one of the allowed symmetry combination, i.e. $L_e-3L_\mu+L_\tau$
from category B for illustration, and 
subject it to constraints from neutral meson mixing data ($K$--$\overline{K}$,
$B_d$--$\overline{B_d}$, $B_s$--$\overline{B_s}$)~\cite{utfit}, global fits on $b \to s \ell \ell$ and $b \to s \gamma$ data~\cite{globalfit4},
and direct detection limits on $Z^\prime$ searches from ATLAS in the di-muon channel~\cite{ATLAS}.
The allowed parameter space
in $\left(g_{Z^\prime},M_{Z^\prime}\right)$ plane is shown in the left panel of fig.~\ref{fig:combinedconstraints}.
The analysis for all the symmetries may be seen in~\cite{ourpaper}. We also show the reach for detecting such 
a $Z^\prime$ with $g_{Z^\prime}=0.36$ at the 13 TeV LHC in the right panel of fig.~\ref{fig:combinedconstraints}. 
\section{Results }
We have arrived at a class of $U(1)_X$ models which can explain
flavour anomalies and neutrino oscillation data simultaneously in a 
bottom-up approach.
A total of nine symmetry combinations are shortlisted demanding theoretical consistencies 
and experimental constraints. An additional Higgs doublet
is introduced to generate the quark mixings. Three right handed neutrinos 
and an additional scalar
singlet are introduced to 
explain neutrino mass and mixings in Type-I seesaw framework.
The  parameters may be chosen such that the additional scalars and
the right handed neutrinos are decoupled from the 
theory. Hence effectively at TeV energies, our model is described by the SM and an 
additional $Z^\prime$. We determine the allowed parameter space in the plane of
($g_{Z^\prime}$, $M_{Z^\prime}$) and also calculate the $Z^\prime$ detection
reach at the 13 TeV LHC in the di-muon channel.


\end{document}